\RequirePackage[2020-02-02]{latexrelease}
\documentclass[twocolumn]{aastex631}
\usepackage{amssymb}
\usepackage{soul}
\usepackage{natbib}
\usepackage{multirow}
\usepackage{graphicx}
\usepackage{tabularx}
\usepackage{verbatim}
\usepackage{color}
\usepackage{ulem}
\usepackage{subfigure}
\usepackage{booktabs,tikz}
\makeatletter
\global\let\tikz@ensure@dollar@catcode=\relax
\makeatother
\usepackage{urwchancal}

\newcommand\soutm{\bgroup\markoverwith
{\textcolor{black}{\rule[0.5ex]{2pt}{0.8pt}}}\ULon}

\shorttitle{Diffuse Light in Low Mass Haloes}
\shortauthors{Contini et al.}

\begin{document}

\title{Diffuse Light in Milky Way-like Haloes}

\correspondingauthor{Emanuele Contini}
\email{emanuele.contini82@gmail.com}
\author{Emanuele Contini}
\affil{Department of Astronomy and Yonsei University Observatory, Yonsei University, 50 Yonsei-ro, Seodaemun-gu, Seoul 03722, Republic of Korea}
\author{San Han}
\affil{Department of Astronomy and Yonsei University Observatory, Yonsei University, 50 Yonsei-ro, Seodaemun-gu, Seoul 03722, Republic of Korea}
\author{Seyoung Jeon}
\affil{Department of Astronomy and Yonsei University Observatory, Yonsei University, 50 Yonsei-ro, Seodaemun-gu, Seoul 03722, Republic of Korea}
\author{Jinsu Rhee}
\affil{Department of Astronomy and Yonsei University Observatory, Yonsei University, 50 Yonsei-ro, Seodaemun-gu, Seoul 03722, Republic of Korea}
\affil{Korea Astronomy and Space Science Institute, 776, Daedeokdae-ro, Yuseong-gu, Daejeon 34055, Republic of Korea}
\author{Sukyoung K. Yi}
\affil{Department of Astronomy and Yonsei University Observatory, Yonsei University, 50 Yonsei-ro, Seodaemun-gu, Seoul 03722, Republic of Korea}

\begin{abstract} 

We investigate the diffuse light (DL) content of dark matter haloes in the mass range $11.5\leq \log M_{halo}\leq13$, a range that includes also the dark matter halo of the Milky-Way, taking advantage of a state-of-the-art semi-analytic model run on the merger trees extracted from a set of high-resolution cosmological simulations. The fraction of DL in such relatively small haloes is found to progressively decrease from the high to the low mass end, in good agreement with analytic (\citealt{purcell2007}) and numerical results from simulations (\citealt{proctor2023,ahvazi2023}), in good agreement also with the fraction of the DL observed in the Milky Way (\citealt{deason2019}) and M31 (\citealt{harmsen2017}). Haloes with different masses have a different efficiency in producing DL: $\log M_{halo} \simeq 13$ is found to be the characteristic halo mass where the production of DL is the most efficient, while the overall efficiency decreases at both larger (\citealt{contini2024}) and smaller scales (this work). The DL content in this range of halo mass is the result of stellar stripping due to tidal interaction between satellites and its host (95\%) and mergers between satellites and the central galaxy (5\%), with pre-processed material, sub-channel of mergers and stripping and so already included in the 100\%, that contributes no more than 8\% on average. The halo concentration is the main driver of the DL formation: more concentrated haloes have higher DL fractions that come from stripping of more massive satellites in the high halo mass end, while dwarfs contribute mostly in the low halo mass end.
\end{abstract}


\section{Introduction}
\label{sec:intro}

The diffuse light (DL), also known as intragroup/cluster light (ICL), is an important component in galaxy groups and clusters (\citealt{contini2021,montes2022,arnaboldi2022} and references therein). On mass scales larger than $\log M_{halo} \geq 13$, it can account for up to 40\% (e.g., \citealt{furnell2021}), or even more in extreme cases (\citealt{tang2018}), of the total light within the virial radius of dark matter (DM) haloes. This diffuse component is made of stars that are not bound to any galaxy, but feel only the potential well of the DM halo, and it is believed to form through tidal stripping of satellite galaxies (e.g., \citealt{contini2014}) and mergers (\citealt{murante2007}) between them and the brightest cluster galaxy (BCG), at epochs that are relatively recent, later than $z\sim 1$ (\citealt{murante2007,contini2014,contini2024}). However, on group scales or below, the disruption or dwarf galaxies (e.g., \citealt{raj2020}) and pre-processed material (e.g., \citealt{ragusa2023}) formed elsewhere and later accreted, become important channels for the formation of the DL.

While a substantial amount of work has been done in order to investigate the ICL and its properties in group/cluster size haloes,  $13\leq \log M_{halo}\leq 15$ (\citealt{burke2015,iodice2017,morishita2017,groenewald2017,demaio2018,montes2018,zhang2019,demaio2020,spavone2020,ragusa2022,ragusa2023,joo2023}, just to quote some of the most recent observational works), only very few authors (e.g., \citealt{purcell2007}, and very recently \citealt{ahvazi2023,proctor2023}) have focused on much smaller scales, as, e.g., Milky Way-like haloes (\citealt{harmsen2017,deason2019}). Understanding the properties of the DL in these haloes, as well as the mechanisms that bring to its formation, is of fundamental importance in order to constrain the typical halo mass under which the formation of the DL becomes negligible, i.e., where either the efficiency of the DM potential well is no longer enough, or the haloes lack of the primary ingredient, satellite galaxies orbiting around the central one.

Very recently, in \cite{contini2023,contini2024}, we have investigated the formation mechanisms of the ICL, and the role played by the halo concentration \footnote{The halo concentration is defined, under the assumption that the halo follows a NFW profile (\citealt{nfw1997}), as the ratio between the virial, $R_{200}$, and the scale, $R_s$, radii of the halo.} in driving it, extending our results up to $z=2$. We have found that the bulk of the formation of the ICL is given by stellar stripping of intermediate/massive satellite galaxies, and that the scatter in the ICL fraction-halo mass relation is driven by the concentration of the halo, in a way that more concentrated (and so relaxed/in a more advanced dynamical state) haloes host also a higher fraction of ICL.

In this Letter, we want to address the main relevant points we addressed for groups and clusters, but in a much smaller range in mass, $11.5\leq \log M_{halo}\leq 13.0$. Our main focuses are what mechanism, and with what strength, is mainly responsible for the formation of the DL, what kind of galaxies contribute the most to it, and more importantly, to answer the following question: is there any particular halo mass where the formation of the DL becomes inefficient? We take advantage of a wide sample of haloes extracted from the merger trees of two N-body simulations (\citealt{contini2023}) with a high resolution in mass ($10^7 M_{\odot}/h$), on which our state-of-the-art semi-analytic model (SAM) ran. Our SAM contains the newest prescriptions for the ICL/DL formation (\citealt{contini2018,contini2019,contini2020}), which account for the relaxation processes happening during galaxy mergers (\citealt{monaco2006,murante2007,burke2015,groenewald2017,jimenez-teja2018,jimenez-teja2019,jimenez-teja2023,joo2023}), stellar stripping due to the tidal forces between the potential well of the halo and satellite galaxies (\citealt{rudick2009,martel2012,contini2014,demaio2015,demaio2018,montes2018,yoo2021,chun2023,zhang2023}), and the amount of diffuse light that can be pre-processed elsewhere and later accreted by larger objects (\citealt{mihos2005,sommer-larsen2006,contini2014,ragusa2023}). We make clear here that our implementation for the DL formation includes also the stellar halo observed in nearby galaxies (e.g., \citealt{longobardi2015}), in the sense that the stellar halo is considered to be part of the DL.

The Letter is organized as follows. In Section \ref{sec:methods} we briefly describe the set of simulations used, the sample of haloes we have obtained, and provide a basic description of the prescriptions used by the SAM to model the DL formation. In Section \ref{sec:results} we present our analysis and results, which will be discussed in Section \ref{sec:discussion} followed by our main conclusions. In the remainder of the manuscript, stellar masses are computed by assuming a \cite{chabrier2003} initial mass function, and all masses have already been h-corrected.

\section[]{Methods}
\label{sec:methods}

We utilize two DM only cosmological simulations originally introduced in \cite{contini2023}, belonging to a wider set of six simulations run with {\small GADGET-4} (\citealt{springel2021}). We make use of the two smallest runs, i.e., YS25HR and YS50HR, which have volumes of $(25\, Mpc/h)^3$ and $(50\, Mpc/h)^3$, numbers of particles $512^3$ and $1024^3$, respectively, that together provide a mass resolution of $10^7$ $M_{\odot}/h$ for both \footnote{Both simulations have the same following Planck 2018 cosmology (\citealt{planck2020}): $\Omega_m=0.31$ for the total matter density, $\Omega_{\Lambda}=0.69$ for the cosmological constant, $n_s=0.97$ for the primordial spectral index, $\sigma_8=0.81$ for the power spectrum normalization, and $h=0.68$ for the normalized Hubble parameter.}. We ran our state-of-the-art SAM (details in \citealt{contini2019,contini2023}) on the merger trees extracted from the set of simulations, and selected haloes in the mass range $11.5\leq \log M_{halo}\leq 13.0$. Our final sample of haloes at $z=0$ comprises 1613 objects, each of them containing a central galaxy and at least one satellite with stellar mass above $\log M_* = 8.0$.

In the context of the goals of this Letter, it is important to briefly summarize the implementation of the DL formation in the SAM. The model accounts for two direct channels, stellar stripping and mergers, and an indirect channel that is pre-processing/accretion. Stellar stripping occurs due to the tidal forces acting on satellites orbiting around the central galaxy. If these forces are strong enough such that the tidal radius is contained within the bulge of the galaxy subject to stripping, it is assumed to be totally disrupted. On the other hand, if the tidal radius is in between the physical limit of the disk of the satellite and its bulge, then only the stellar mass in that shell is stripped. In both cases the stellar mass stripped ends up in the DL component associated to the central galaxy, which includes also the stellar halo observed in nearby galaxies. The tidal radius, $R_t$, is the radius at which the gravity of the satellite is no longer able to keep it stable, and it is estimated in the model with the following equation (see \citealt{binney2008}):
$$  R_{t} = \left(\frac{M_{sat}}{3 \cdot M_{halo}}\right)^{1/3} \cdot D \, , $$
where $M_{sat}$ is the satellite mass (stellar mass + cold gas mass), $M_{halo}$ is the DM mass of the halo, and $D$ the satellite distance from the centre of the halo. This equation is directly applied to satellites for which their subhalo went under the resolution of the simulation or has been disrupted. The same equation is applied to satellites still associated with a subhalo, but an extra condition must be satisfied in the first place, i.e.:
$$ R^{DM}_{half} < R^{Disk}_{half} \, ,$$
where $R^{DM}_{half}$ is the half-mass radius of the parent subhalo, and $R^{Disk}_{half}$ the half-mass radius of the galaxy's disk.

During mergers between satellites and the centrals, instead, 20\% of the mass \footnote{The reader can find a full discussion of  the percentage chosen in \cite{contini2014}.  Taking advantage of control simulations of galaxy groups (\citealt{villalobos2012}) we derived that the typical fraction of stellar mass that gets unbound during mergers, both minor and major, peaks on 0.2. In later works (e.g. \citealt{contini2018,contini2019}) we confirmed the validity of the assumption on larger scales.} of the merging satellite goes to the DL component, while the rest of it effectively merge with the central galaxy. Pre-processing accounts for DL formed elsewhere through either stripping or mergers, but accreted by a given halo during its assembly. For further details of the implementations we refer the reader to the papers quoted above.

\section{Results}
\label{sec:results}

\begin{figure*}
\centering
\includegraphics[width=0.9\textwidth]{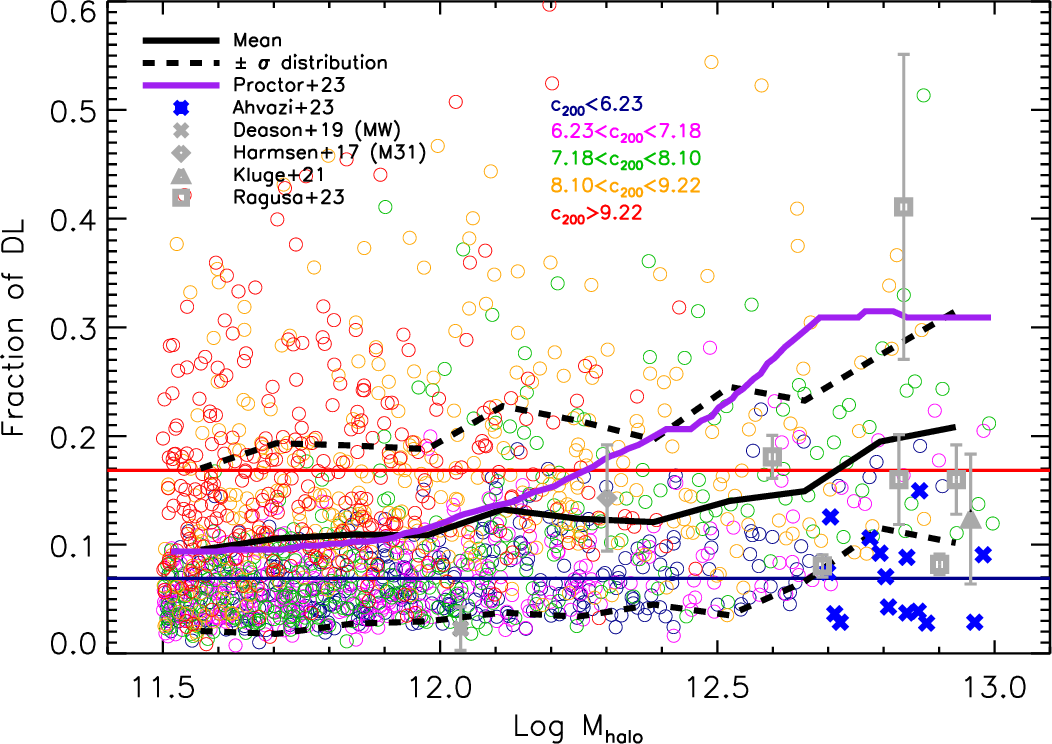}
\caption{Fraction of DL as a function of halo mass for haloes in different ranges of concentrations as shown in the legend, while the black solid and dashed lines indicate the mean and $\pm\sigma$ scatter of the distribution. Our results are compared with the predictions from \citealt{proctor2023} (Eagle Project), from \citealt{ahvazi2023} (TNG50), and several observational measurements (\citealt{deason2019,harmsen2017,ragusa2023,kluge2021}). The DL fraction weakly increases towards larger haloes from about 0.1 in the low mass end to 0.2 in the high mass end, but the distribution also shows a $\sim\pm 0.1$ scatter along the whole mass range. Considering the typical fractions found in groups and clusters (\citealt{contini2023,contini2024}), overall the DL fraction decreases quite importantly towards low mass haloes, potentially becoming negligible on scales below those investigated here. It is important to highlight the role played by the concentration even in these scales: more concentrated, and so relaxed haloes, tend to host a higher fraction of DL, as shown by the red (average DL fraction in the highest bin in concentration) and blue (average DL fraction in the lowest bin in concentration) lines, respectively.}
\label{fig:iclhalomass}
\end{figure*}

\begin{figure}
\centering
\includegraphics[width=0.47\textwidth]{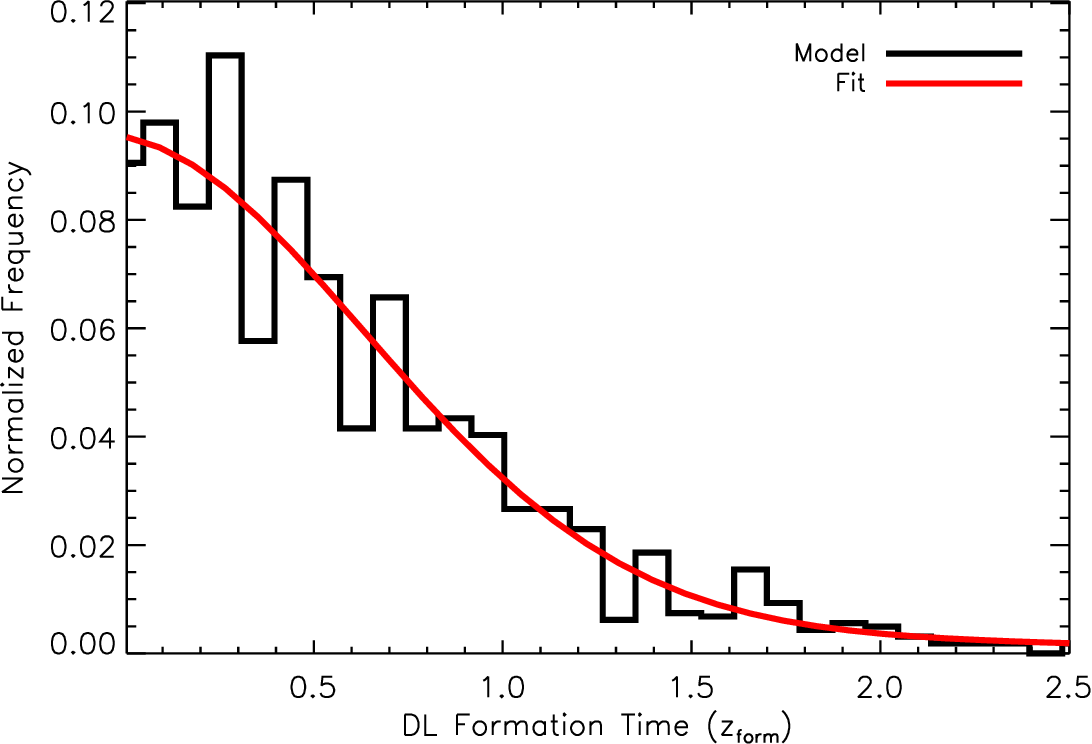}
\caption{Distribution of the DL formation time in our sample of haloes indicated by the black histogramm, and its fit indicated by the red line. The formation time has been defined as the redshift when 50\% of the DL at $z=0$ was already formed. Exactly 50\% of the haloes have formation times $z_{form}\leq 0.5$, and around a third of them (32\%) have $z_{form} \leq 0.3$. These numbers, along with a simple comparison with the distribution found in Fig. 7 of \cite{contini2024} for groups and clusters, point out that these haloes are still in the process of their formation, and so is the DL they host.}
\label{fig:histo}
\end{figure}

\begin{figure}
\begin{center}
\begin{tabular}{cc}
\includegraphics[scale=.46]{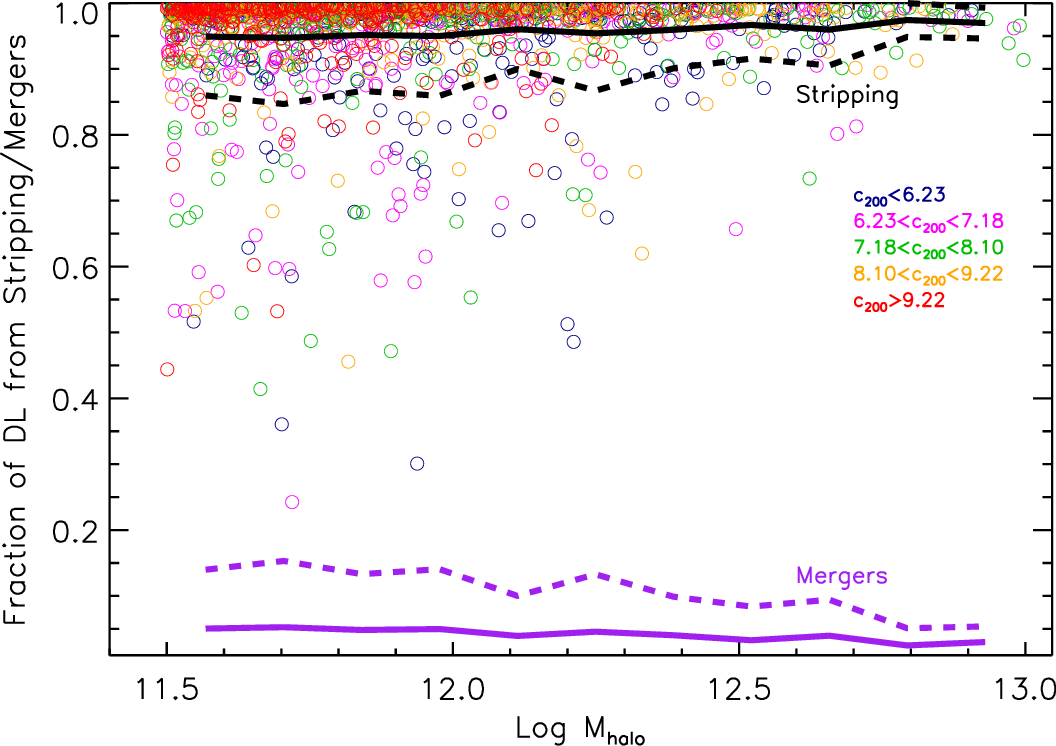} \\
\includegraphics[scale=.46]{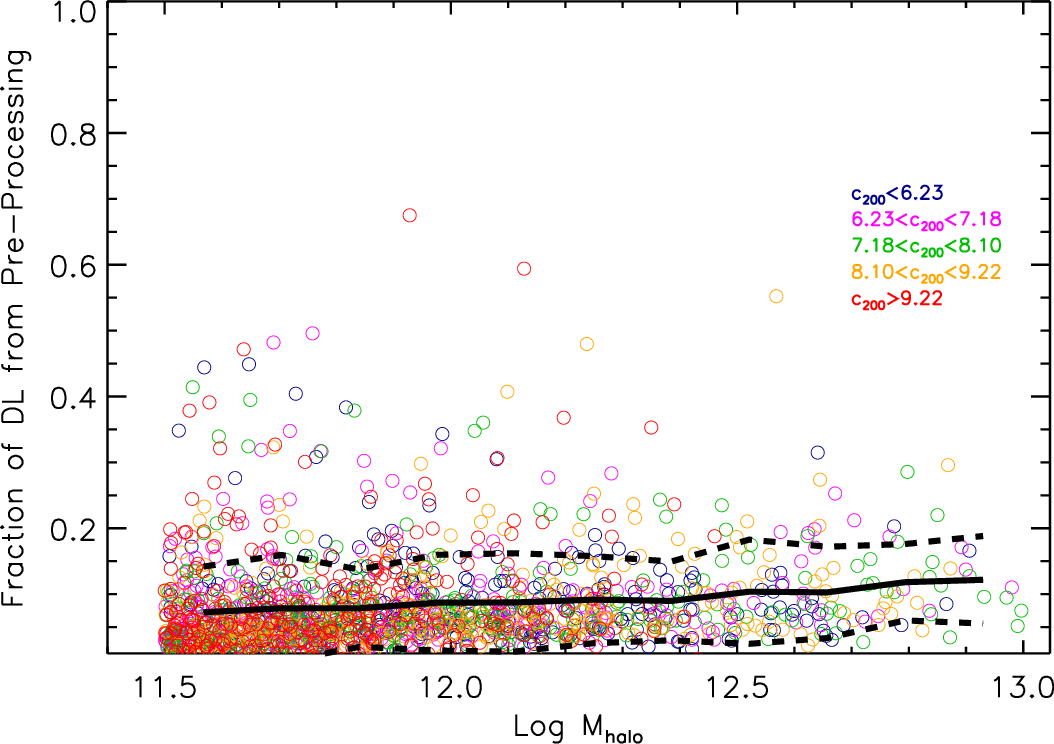} \\
\end{tabular}
\caption{Top panel: fraction of DL coming from the stellar stripping (black solid and dashed lines), and from the merger (purple solid and dashed lines) channels. In both cases the solid and dashed lines indicate the mean and $\pm\sigma$ scatter of the distributions, while the circles with different colors represent haloes in different ranges of concentrantions in the case of the stripping channel, as shown in the legend. Similarly to what found in \cite{contini2024} for groups and clusters, stellar stripping is the major responsible (even more in more concentrated haloes) for the bulk of the DL (95\% on average), independently on the halo mass, while mergers play a marginal role (5\% on average). Bottom panel: the same as in the top panel, for the case of the DL coming from pre-processing/accretion. Differently from their much larger counterparts, these haloes accrete, on average, 8\% of the 100\% of DL given by mergers and stripping, a number that highlights the fact that these haloes are still in the process of their formation and did not accrete much material from outside, having had so far a few minor or no relevant major mergers.}
\label{fig:iclcontr_hm}
\end{center}
\end{figure}

\begin{figure}
\centering
\includegraphics[width=0.47\textwidth]{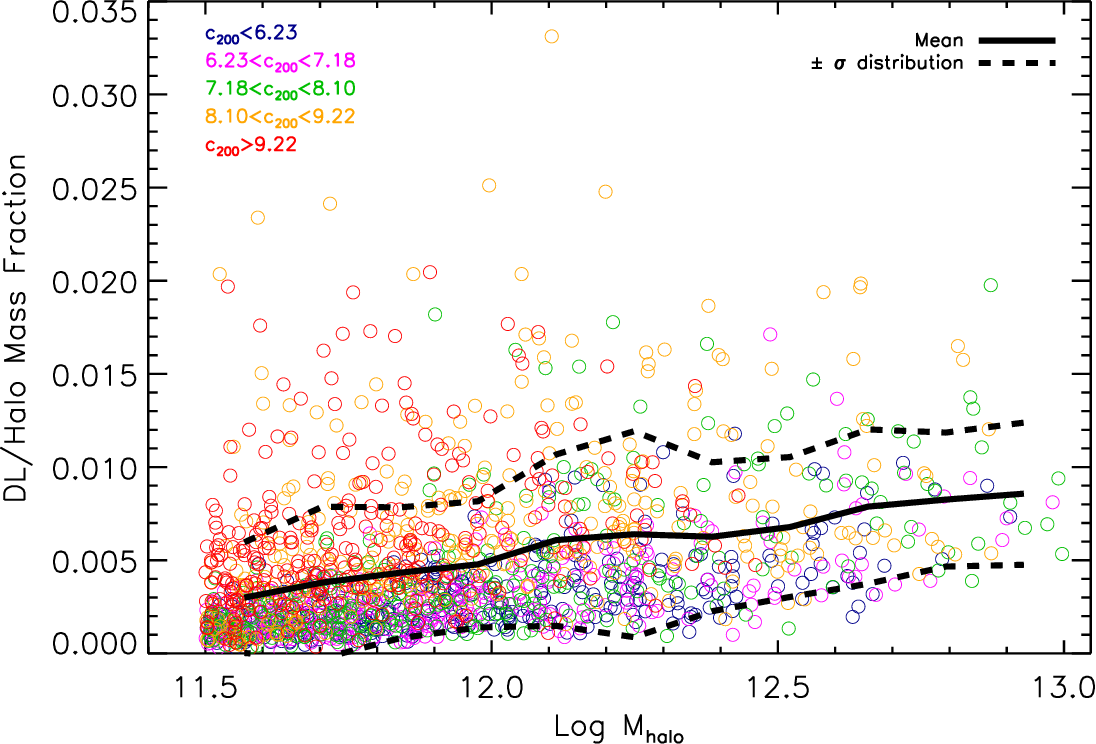}
\caption{Ratio between the amount of DL and the halo mass of its host, as a function of the latter. Circles with different colors represent haloes in different ranges of concentration, as indicated in the legend, while the black solid and dashed lines represent the mean and $\pm\sigma$ scatter of the distribution. This ratio, which can be considered as a sort of efficiency of haloes in producing DL, is an increasing function of halo mass, from around 0.3\% (lowest halo mass) to 0.8\% (highest halo mass). Putting together this result and its equivalent found in \cite{contini2024} for groups and cluster, where the ratio is a weak decreasing function of halo mass, the net conclusion is that haloes with $\log M_{halo}\sim 13$ are the most efficient ones in producing DL. Once more, it is important to underline the role of the concentrantion: independently on the halo mass, more concentrated haloes are more efficient.}
\label{fig:iclfhm}
\end{figure}

\begin{figure}
\begin{center}
\begin{tabular}{cc}
\includegraphics[width=0.445\textwidth]{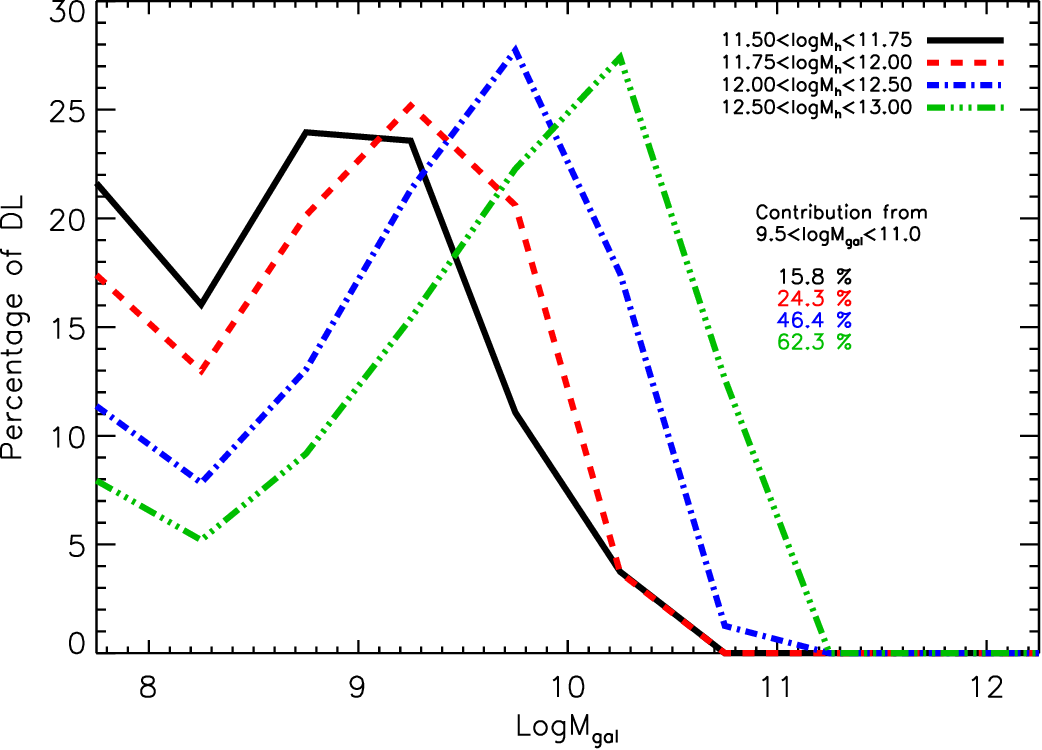} \\
\includegraphics[width=0.46\textwidth]{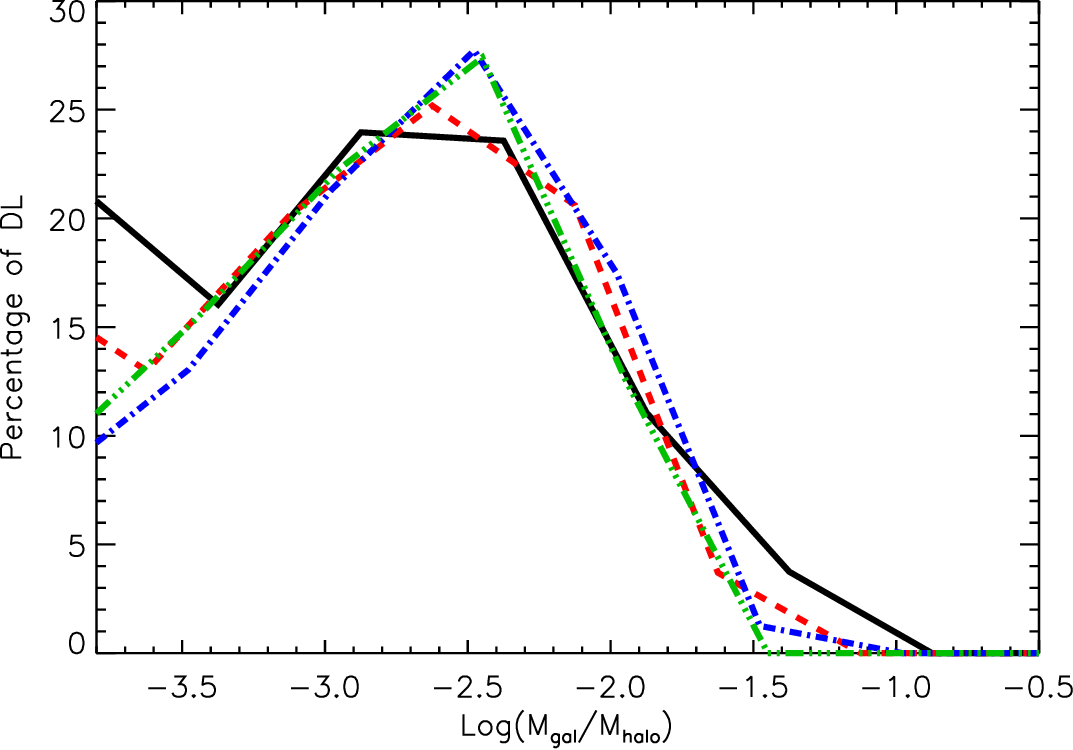} \\
\end{tabular}
\caption{Percentage of DL as a function of the stellar mass of the galaxies that contribute to it (top panel), for haloes in four different ranges of mass as indicated in the legend, and normalized to the halo mass in each range (bottom panel). The peak of the distributions moves towards the left (lower stellar mass) as the range in halo mass decreases, which means that the galaxies contributing to the DL are smaller and smaller as the halo mass decreases. If we focus on the contribution given by relative intermediate/massive galaxies ($9.5<\log M_{*}<11$), we find that their contribution to the DL increases rapidly with increasing halo mass range, from around 16\% in the lowest halo masses, to 62\% in the largest ones. This result is in line with what we found in \cite{contini2014}, i.e., that intermediate/massive galaxies contribute the most to the DL. However, when the halo mass decreases substantially, most of the contribution is given by galaxies with stellar mass lower or much lower than $\log M_{*}\sim 9.5$. Once we scale the x-axis to the halo mass of host galaxies (bottom panel), the four distributions look similar and the peak lies between $0.001<\log (M_{gal}/M_{halo})<0.003$.}
\label{fig:iclmgal}
\end{center}
\end{figure}

\begin{figure}
\begin{center}
\begin{tabular}{cc}
\includegraphics[width=0.45\textwidth]{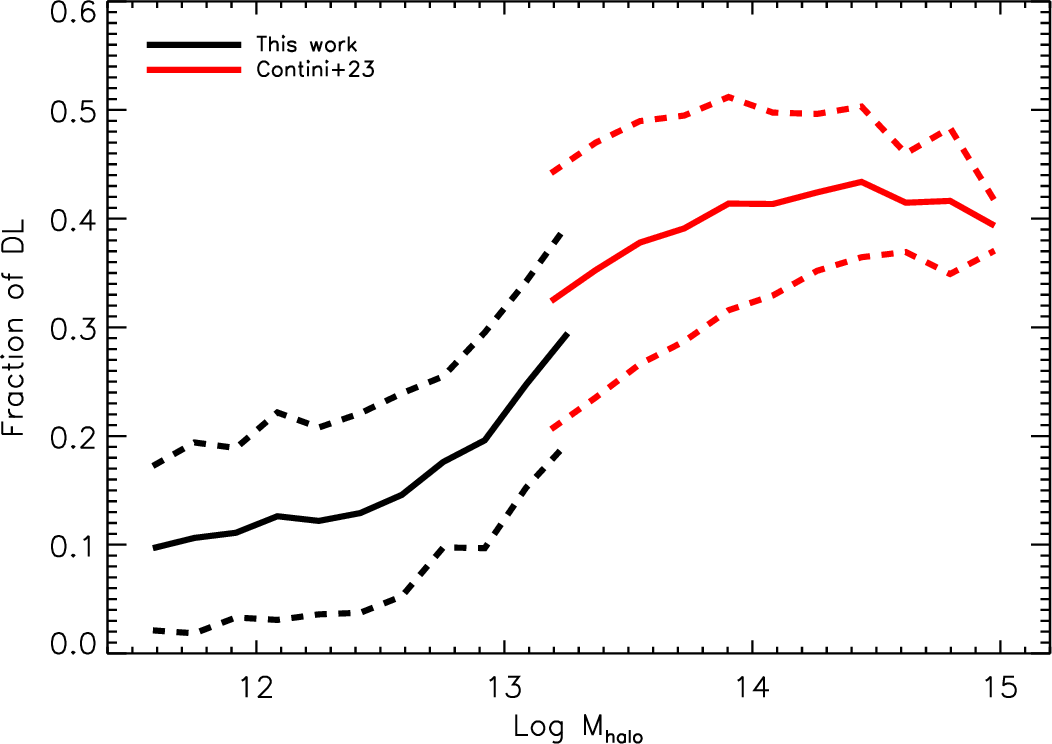} \\
\includegraphics[width=0.46\textwidth]{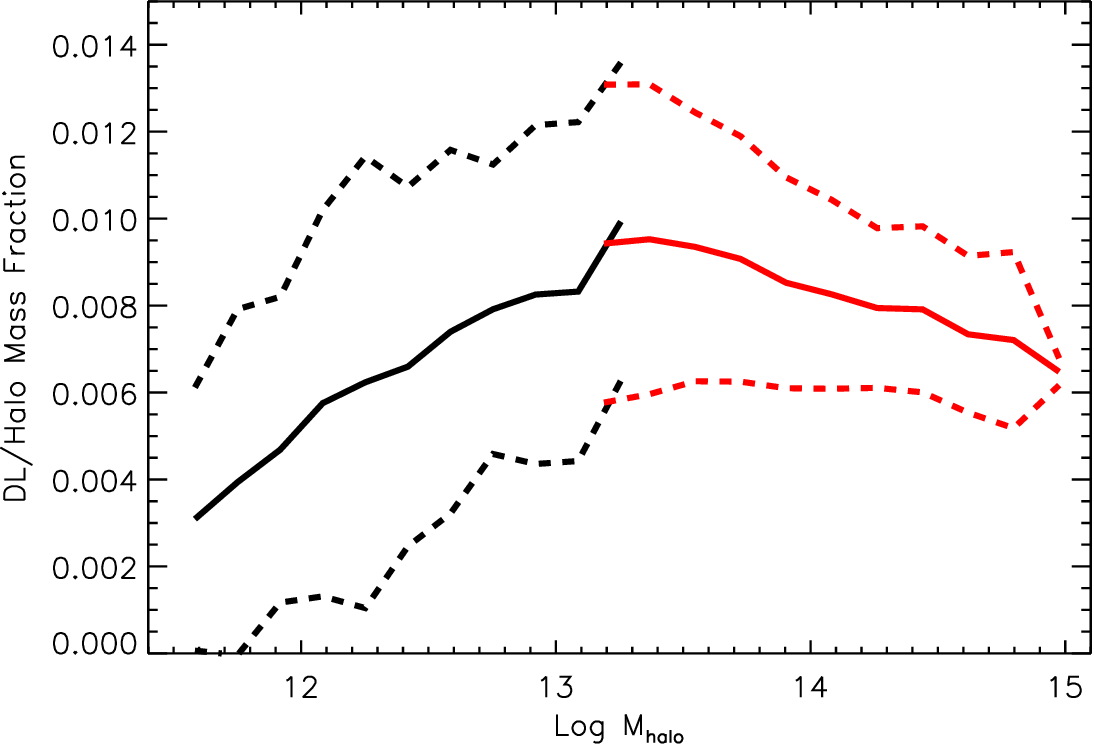} \\
\end{tabular}
\caption{Full comparison between our results in \cite{contini2023,contini2024} and this work. In the top panel we plot the DL fraction as a function of halo mass in the range $11.5<\log M_{halo}<15.0$ by connecting the relation found in this work (black lines) and that in \citealt{contini2023} (red lines). Similarly, in the bottom panel we put together the relations between the DL/halo mass ratio and halo mass found here and in \cite{contini2024}. The DL fraction increases from very small
scales to large clusters, while the DL/halo mass ratio, which we consider as the efficiency of haloes to form DL, increases from very small scales up to the peak in $\log M_{halo}\sim 13$, to decrease again towards large clusters. However, two important caveats must be noted, and the details are specified in the text.}
\label{fig:gen_pic}
\end{center}
\end{figure}

The main goal of this Letter is to understand what is the typical halo mass below which the production of DL starts to become inefficient, and smaller haloes than that typical mass would have progressively lower DL fraction which eventually becomes negligible. This implies also a series of results that we are going to show through the analysis below. In order to achieve our goal, we first start by showing in Figure \ref{fig:iclhalomass} the fraction of DL as a function of halo mass, and compare the predictions of our model (black lines and circles with different colors) with the results of the Eagle project (\citealt{proctor2023}, purple line), those of TNG50 (\citealt{ahvazi2023}, blue crosses), and several observed fractions from different authors (\citealt{deason2019,harmsen2017,ragusa2023,kluge2021}, grey symbols). The plot shows that the predictions of our model are in better agreement with observations than the other two numerical predictions, and it does agree with the Eagle project on very small scales, up to $\log M_{halo} \sim 12.5$ if we consider the scatter (but they start to diverge from $\log M_{halo} \sim 12.1$). Regardless the comparisons with other works, the model predicts that the DL fraction becomes smaller soon after $\log M_{halo} <12.6$ and converges fast to $\sim 0.1$ towards the low-mass end of our sample. Remarkably, as found in \cite{contini2023,contini2024}, even on these scales the role of the halo concentration is very important: more concentrated and relaxed haloes tend to host higher fractions of DL,
as shown by the red (average DL fraction in the highest bin in concentration) and blue (average DL fraction in the lowest bin in concentration) lines, respectively.

In Figure \ref{fig:histo} we show the distribution of the DL formation time (black line) and its fit (red line), defined as the redshift when 50\% of the DL at $z=0$ was already formed.
Half of the haloes in the sample have formation times $z_{form}\leq 0.5$, and one third of them have $z_{form} \leq 0.3$. These numbers are very important because they remark the fact that most of the haloes in the sample are still in the process of their formation, and considering that the DL forms in parallel with its host (\citealt{contini2024}), so is the DL they
host. What mechanisms are causing the formation of the DL? We address this point in Figure \ref{fig:iclcontr_hm}, where we plot the contribution to the DL from stellar stripping and mergers in the top panel, and from the pre-processing channel in the bottom panel. Very similar to what found in \cite{contini2024}, stellar stripping dominates over all the halo mass range investigated, being responsible for 95\% of the total amount of DL and leaving to mergers only the remaining 5\%. Contrary to groups and clusters, pre-processing is less important, accounting for no more than 8\% (on average) of the total DL due to mergers and stripping together (we remind the reader that pre-processing is a sub-channel of mergers and stripping). As stated above, these haloes are mostly in the process of their formation, i.e., no much DL should be expected from accretion if the host has expericend only a few minor or no relevant major mergers.

Finally, we present in Figure \ref{fig:iclfhm} the ratio between the DL and its host mass, as a function of the latter. In \cite{contini2024}, we associated this ratio to the efficiency
of the hosts in producing DL, and found that it is a weak decreasing function of halo mass, from an average 0.9\% in small groups ($\log M_{halo}\sim 13$) to an average 0.5\% in large clusters ($\log M_{halo}\sim 15$), although with a wide scatter. In the halo mass range investigated here, we find the opposite trend, i.e., the ratio decreases from about 0.8\% (highest halo mass) to 0.3\% (lowest halo mass). Putting the pictures altogether, it follows that the characteristic halo mass (at $z=0$) most efficient (on average) in producing DL is $\log M_{halo} \simeq 13.0$, and the DL it hosts is mainly coming from the stripping channel (but see next section for an important caveat on this point).

Most of the DL forms through stellar stripping, and we already know (\citealt{contini2014}) that intermediate/massive galaxies are responsible for the bulk of it. The last question we want to answer, in order to complete the analysis is: are intermediate/massive galaxies still the main contributors? The answer to this question is in Figure \ref{fig:iclmgal}, where we plot the percentage of DL that comes from stellar stripping of galaxies having a given stellar mass (top panel), for haloes in different mass ranges as indicated in the legend of the Figure. We can clearly see that the peak of the distributions progressively decreases, from $\log M_* \sim 10.3$ in the highest halo mass range, to $\log M_* \sim 9.0$ in the lowest. The contribution of galaxies in the stellar mass range $9.5<\log M_* < 11.0$ increases from 16\% to 62\% in the same ranges, implying that while intermediate/massive galaxies still contribute remarkably in the formation of the DL in haloes with mass close to the most efficient in producing DL, their contribution becomes almost negligible in very small haloes, but still important in Milky Way-like haloes, between 24\% and 46\%. It must be noted, however, that the different distributions are a natural consequence of the fact that larger haloes can host larger satellites, which are also those more subject to stripping. Once the x-axis in Figure \ref{fig:iclmgal} is scaled to the mass of the host (bottom panel), the main differences in the distributions disappear (see the discussion on single case studies in the next section) and the peak lies between $0.001<\log (M_{gal}/M_{halo})<0.003$.

\section{Discussion and Conclusion}
\label{sec:discussion}

The first theoretical prediction of the DL fraction in such small scales is dated to almost two decades ago, when \cite{purcell2007} showed that their analytic model predicts an increasing fraction of DL for haloes in the range $10.5<\log M_{halo}<15$, negligible below $\log M_{halo} \sim 12$, but around 30\% on cluster scales. Their work gave the first hint for a significant different amount of DL in haloes much smaller than those usually probed in observations ($\log M_{halo}>12.5$). However, in the past few years there have been the first measurements of the DL in the Milky Way (\citealt{deason2019}) and M31 (\citealt{harmsen2017}), and given the improvements in the spatial and mass resolutions of hydro-simulations, also the first theoretical predictions (e.g., \citealt{proctor2023,ahvazi2023}). Our model has never been tested on such small scales, and this Letter brings the first results on the formation of the DL in haloes below the typical halo masses usually probed.

The first key-point of this work is that the fraction of DL in Milky Way-like haloes is in good agreement with the observations of the stellar halo of the Milky Way by \cite{deason2019}, and that of M31 by \cite{harmsen2017} (Fig.\ref{fig:iclhalomass}). Not only, our model predicts fractions that are in good agreement with the observations by \cite{ragusa2023} and \cite{kluge2021} in that halo mass range. Overall, the model agrees very well with the predictions of the Eagle project (\citealt{proctor2023}) up to $\log M_{halo} \sim 12.1$, while those of TNG50 (\citealt{ahvazi2023}) are biased-low with respect to ours, and predicts DL fractions that decrease quite importantly towards low mass haloes. The disagreement between Eagle and TNG50 simulations, in
our opinion, has to be taken with caution. In fact, the two simulations have different mass resolutions and more importantly, the authors adopt different definitions for the DL. On one hand, \cite{proctor2023} use Gaussian Mixture Models to separate bulge, disk and DL. On the other hand, \cite{ahvazi2023} define DL the amount of stars that are bound only to the potential well of the main halo in the range $0.15R_{200}<r<R_{200}$.

In our sample, most of the haloes are still in the process of their formation, and so is the DL. Haloes that are more relaxed, that is, more concentrated, tend to host higher fractions of DL, due to the stronger tidal forces acting on satellite galaxies that are subject to stellar stripping. We have conducted a thorough
study on single cases among our sample in order to check whether or not single episodes of stripping of intermediate/massive galaxies might play an important role. By selecting haloes in narrow ranges of mass in the low mass end, we have found that haloes with higher fractions of DL with respect to the average, most of the times form their DL with a few stellar stripping events, sometimes with a significant stripping of intermediate/massive satellites. On the other hand, haloes with lower DL fractions with respect to the average are instead characterized by a larger number of surviving satellites which did not meet yet the conditions for stripping. This implies that the level of relaxation in such small haloes, which (we remind the reader) are still assembling, makes a real difference in the production of DL. In a few words, a few single episodes of stripping can remarkably move the fraction of DL from below to above the average value.

It is not surprising that pre-processing and mergers are not contributing much to the formation of the DL on these scales. As said, these haloes are in the process of their formation and
while mergers between haloes bring new material that can be then subject to stripping, they also account for the pre-processed DL. Another thorough analysis of single cases led us to the conclusion that the scatter in the bottom panel of Figure \ref{fig:iclcontr_hm} can be explained by the number of mergers experienced since the DL started to form. Haloes that have typically experienced a few mergers hosts a larger fraction of pre-processed DL, but still marginal if compared to their more massive counterparts in the group and cluster scales.

Another important aspect of the analysis is that stellar stripping dominates the production of the DL, at all scales probed and even more significantly than what found in \cite{contini2024} for groups and clusters. This channel accounts for an average of 95\% of the DL, leaving to mergers just a marginal role. However, as noticed in \cite{contini2024}, these fractions must be taken with a pinch of salt because they are significantly dependent on the particular definition used for mergers. In studies such as \cite{murante2007}, or Joo et al. 2024 (in prep.), the definition of a merger is more relaxed than ours, leading to a scenario where most of the DL in our stripping channel actually belongs to the merger channel in theirs. This was
demonstrated and quantified in \cite{contini2018} where, by relaxing our definition of the stripping channel, we found the same percentages invoked by \cite{murante2007}. 

Most of the DL coming from stellar stripping, which represent almost the total, is stripped by intermediate/massive galaxies even in some haloes of the sample probed in this work. However, this contribution becomes less and less important when the halo mass decreases, and at Milky Way/M31 like halo scales it accounts for around 30-35\% (between 24\% and 46\%), to eventually drop to 16\% at lower masses. The low-mass scales are characterized by haloes that produce their DL by stripping dwarf galaxies, with episodic events of stripping of more massive satellites. This has an important effect on the production of DL in such small haloes, because it will eventually become inefficient and the fractional budget of the DL is expected to drop to zero in halo masses lower than those probed here, as also predicted by the analytic model of \cite{purcell2007}.

In Figure \ref{fig:gen_pic}, the main results of the analysis done here are put together with the results found in \cite{contini2023,contini2024}. In the top panel we compare the DL fraction found in the range investigated here with that in \cite{contini2023}, and in the bottom panel we compare the efficiency in producing DL as a function of halo mass connecting the results of this work and those found in \cite{contini2024}. The DL fraction nicely increases from very small haloes to large clusters, and so does the efficiency up to $\log M_{halo}\sim 13$, to decrease again towards large clusters. It must be noted, however, that there is no observational evindence yet for a rise of the DL fraction towards larger haloes, and this prediction of the model has to be taken with caution until we collect more observed data that can confirm it. The shape of the relation between the efficiency and the halo mass puts a constraint on the typical halo mass where the production of DL becomes the most efficient. This happens in haloes with $\log M_{halo} \sim 13$, where the ratio between DL and halo mass is the highest, around 0.09\%. In every halo mass scale the concentration is driving the formation of the DL, but while in large haloes the main material (satellite galaxies subject to stripping) is always present, on scales lower than the characteristic mass where the efficiency decreases again, the ingredient needed to form DL starts to lack, and the mechanism of stripping becomes inefficient simply because of the lack of satellites that can be subject to stellar stripping.

However, there are two caveats that must be noted regarding the comparisons shown in Figure \ref{fig:gen_pic}. In order to match the range investigated in \cite{contini2023} and \cite{contini2024}, we increased the halo mass range probed here to $\log M_{halo} \sim 13.3$ of the set of simulations used in this work, while the other range remains untouched. The simulations used in the former studies have a resolution that is, on average, slightly more than an order of magnitude lower, which leads to predict a DL fraction somewhat higher in the same halo mass scale (see \citealt{contini2014} for the dependence on the resolution). Nevertheless, the comparison is safe considering the statistics in the halo mass range involved. The other caveat is that we have poor statistics in this set of simulations (high resolution) on that scale, but again compensated by the large statistics from the other set (lower resolution). All in all, resolution and statistics around $\log M_{halo}\sim 13-13.2$ do not have a net effect, but mentioning them is mandatory.

To summarize, the formation of the DL in haloes with mass lower than $\log M_{halo} =13$, which includes also Milky Way-like haloes, is driven by the halo concentration. More concentrated and relaxed haloes host a higher fraction of DL that forms from stellar stripping of satellite galaxies, with more massive satellites contributing mostly in the high mass end and dwarfs in the low mass end. However, on these scales most of the haloes are actually still on the way of their assembly, and so is the DL they host. It is reasonable to think that a thorough analysis on scales lower than those probed here would reveal a drop in the DL fraction, simply because the mechanism responsible for the DL formation is inhibited by the lack of sufficiently massive satellite galaxies.


\section*{Acknowledgements}
E.C. and S.K.Y. acknowledge support from the Korean National Research Foundation (2020R1A2C3003769). E.C. and S.J. acknowledge support from the Korean National Research Foundation (RS-2023-00241934). All the authors are supported by the Korean National Research Foundation (2022R1A6A1A03053472). J.R. was supported by the KASI-Yonsei Postdoctoral Fellowship and by the Korea Astronomy and Space Science Institute under the R\&D program (Project No. 2023-1-830-00), supervised by the Ministry of Science and ICT.


\bibliography{paper}{}
\bibliographystyle{aasjournal}

\end{document}